\documentclass[12pt]{article}
\usepackage{fullpage}
\usepackage{amsmath}
\usepackage{amsfonts}
\usepackage{amssymb}
\usepackage{graphicx}
\usepackage{subfigure}
\usepackage{psfrag}
\usepackage{rotating}
\usepackage{lscape}
\usepackage{geometry}
\usepackage{rotating}
\usepackage{url}
\usepackage{float}

\numberwithin{equation}{section}

\allowdisplaybreaks[3]

\date{\today}

\def\bb{{b}}
\def\hb{\hat{b}}
\def\hbb{\hat{b}}

\def\rE{{\mathrm E}}

\def\rP{{\mathrm P}}

\def\hQ{\hat{Q}}

\def\hS{\hat{S}}

\def\ts{\tilde{s}}

\def\hatt{\hat{t}}
\def\tt{\tilde{t}}

\def\bu{u}

\def\bv{{\boldsymbol v}}

\def\hV{\hat{V}}

\def\bx{x}

\def\by{y}

\def\hz{\hat{z}}
\def\tz{\tilde{z}}

\def\eps{\epsilon}
\def\veps{\varepsilon}
\def\bveps{{\varepsilon}}

\def\argmin{\mathop{\rm argmin}}
\def\real{\mathop{{\rm I}\kern-.2em\hbox{\rm R}}\nolimits}

\def\hbeta{\hat{\beta}}

\def\trace{\mbox{trace}}

\def\eps{\epsilon}

\def\bbeta{\beta}
\def\hbbeta{\hat{\beta}}

\def\tbeta{\tilde{\beta}}
\def\tbbeta{\tilde{\beta}}

\def\gam{\gamma}

\def\lam{\lambda}

\def\hlam{\hat{\lambda}}

\def\drho{\dot{\rho}}

\def\sig{\sigma}
\def\hsigma{\hat{\sigma}}

\def\sgn{\hbox{sgn}}
\def\Var{\hbox{Var}}
\def\Cov{\hbox{Cov}}

\def\Fdp{\textrm{Fdp}}

\def\Fmp{\textrm{Fmp}}

\newcommand{\cov}{\textrm{Cov}}
\newcommand{\Cor}{\textrm{Corr}}

\def\T{{ \mathrm{\scriptscriptstyle T} }}

\newtheorem{theorem}{Theorem}

\begin{document}

\begin{center}
\textbf{\large 
Semi-Penalized Inference with Direct False Discovery Rate Control in High-Dimensions}

Jian Huang$^{1}$, Shuangge Ma$^{2}$, Cun-Hui Zhang$^{3}$ and Yong Zhou$^{4}$
\end{center}

\noindent
1. Department of Statistics and Actuarial Science, and Department of Biostatistics, University of Iowa, 
Iowa City, Iowa 52242, U.S.A.

\noindent
2. Department of Biostatistics, School of Public Health, Yale University,
New Haven, Connecticut 06520, U.S.A.

\noindent
3. Department of Statistics and Biostatistics, Rutgers University, Piscataway, New Jersey 08854, U.S.A.

\noindent
4. School of Statistics and Management, Shanghai University of
Finance and Economics, Shanghai 200433, China


\medskip\noindent
\begin{center}
\textbf{Abstract}
\end{center}
We propose a new method, semi-penalized inference with direct false discovery rate control (SPIDR), for variable selection and confidence interval construction in high-dimensional linear regression.
SPIDR first uses a semi-penalized approach to constructing estimators of the regression coefficients.  We show that the SPIDR estimator is ideal in the sense that it equals an ideal least squares
estimator with high probability under a sparsity and other suitable conditions. Consequently, the SPIDR estimator is asymptotically normal.   Based on this distributional result, SPIDR determines the selection rule by directly controlling false discovery rate. This provides an explicit assessment of the selection error. This also naturally leads to confidence intervals for the selected coefficients with a proper confidence statement. We conduct simulation studies to evaluate its finite sample performance and demonstrate its application on a breast cancer gene expression data set. Our simulation studies and data example suggest that SPIDR is a useful method for high-dimensional statistical inference in practice.

\medskip\noindent
\textit{Some key words.} Confidence interval; Selection Error; Concave penalty; Variable selection; Sparsity; Stickiness.

\section{Introduction}

Consider the linear regression model
\begin{equation}
\label{LinMod1}
\by = \sum_{j=1}^p \bx_j\beta_j + \bveps,
\end{equation}
where $\by=(y_1, \ldots, y_n)'$  is a vector of response variables,
$\bx_j =(x_{1j}, \ldots, x_{nj})'$ is the $j$th vector of predictors, $\beta_j$ is the $j$th regression coefficient and $\bveps=(\veps_1, \ldots, \veps_n)'$ is a vector of error terms. Here $p$ is the number of predictors and $n$ is the sample size.
Let $S=\{j: |\beta_j|>0, 1\le j\le p\}$ be the support of $\bbeta$.
We are interested in the high-dimensional case where $p \gg n$ and the model is sparse in the sense that the cardinality of $S$ is
small relative to $n$.  We propose a new approach for variable selection and confidence interval construction based on semi-penalized
inference with direct false discovery rate control . For brevity, we shall simply refer to the proposed methodology as SPIDR.

There is now a substantial body of work on penalized methods for variable selection. Several important penalty functions have been introduced.
Examples include the least absolute shrinkage and selection operator (Lasso) or the $\ell_1$ penalty (Tibshirani (1996)), the smoothly clipped absolute deviation (SCAD) penalty (Fan and Li (2000)), and the minimum concave penalty (MCP, Zhang (2010)).
A common feature of these penalties is that they are capable of producing exact zero solutions, which automatically leads to variable selection.
The penalized methods also enjoy many attractive theoretical properties
concerning the selection, estimation and prediction in sparse, $p \gg n$ settings, including the asymptotic oracle property under certain conditions.  But they do not provide a computable error assessment of the selection results in finite sample situations.
The literature on this topic has grown too vast to be adequately summarized here, so we refer to the book by B\"{u}hlmann and van de Geer (2011) and the references therein for the results on convex selection,
and Fan and Li (2000), Fan and Lv (2011), Zhang (2010) and Zhang and Zhang (2013) and the references therein for the results on concave selection.

On a different front in the area of high-dimensional data analysis,
many researchers have considered the problem of large scale hypothesis testing. In particular, since the appearance of the seminal paper of Benjamini and Hochberg (1995), false discovery rate (FDR) has become a widely accepted error measure in scientific investigations involving a large number of hypotheses, such as genomic studies with data from array-based technology (Storey and Tibshirani (2003)). In recent years, there has been a growing interest in applying the ideas of FDR in the estimation of sparse, high-dimensional models. Abramovich et al. (2006) introduced an FDR-based thresholding approach for estimating a sparse mean vector $\mu \in \real^n$ based on an observation $y\in \real^n$ from a multivariate normal model $N(\mu, \sigma_n^2 I_n)$, where $I_n$ is an $n\times n$ identity matrix and $\sigma_n^2>0$ is assumed to be known for theoretical analysis. They obtained in-depth asymptotic minimaxity results under various sparsity conditions on $\mu$. A key factor that enables the
construction of the FDR-based thresholding rule and theoretical analysis
is the availability of the estimator $y \sim N(\mu, \sigma_n^2 I_n)$. Indeed, their FDR-based thresholding rules are defined using the ordered values of the components in $y$.
Benjamini and Gavrilov (2009) proposed a step-wise forward selection, which tests the coefficients
and adds variables sequentially using a multiple-state FDR correction.
Meinshausen and B\"{u}hlmann (2010) introduced stability selection that uses resampling to evaluate the probability of each variable being selected.  It provides an upper bound for the expected number of falsely selected variables under an exchangeability condition. This approach was further refined by Shah and Samworth (2013). Meinshausen, Meier
and B\"{u}hlmann (2009) used sample splitting to obtain the $p$-values
for the predictors. B\"{u}hlmann (2012) proposed a method
for constructing $p$-values based on ridge estimation with an additional bias correction step in high dimensions.
However, these works did not make an explicit connection with the direct estimation of FDR in the context of variable selection.

A third recent development is on the statistical inference for low-dimensional parameters in high-dimensional models.
Zhang and Zhang (2011) proposed a semiparametric efficient score
approach for constructing confidence intervals of low-dimensional
coefficients in high-dimensional linear models. Van de Geer, B\"{u}hlmann and Ritov (2013) considered the same problem by using an approach
that inverts the optimization conditions for the Lasso solutions.
They extended the work of Zhang and Zhang (2012) to
generalized linear models and problems with convex loss functions.
Javanmard and Montanari (2013) considered the problem of
hypothesis testing in high-dimensional regression using a method similar
to that of Zhang and Zhang (2011).
Belloni, Chernozhukov and Hansen (2012) proposed a two-stage selection
procedure with post-double-selection to estimate a single treatment
effect parameter in a high-dimensional liner model.
These authors did not consider the problem of variable selection or direct FDR control.

In this paper, we formulate the problem of variable selection in the
framework of large scale hypothesis testing based on the semi-penalized
estimators. This enables us to utilize the methods for multiple comparisons to assess the selection error. There are two essential ingredients in SPIDR, the first is the estimation of regression coefficients; the second is selection and confidence interval construction with FDR control.
To study the theoretical properties of the SPIDR estimator, we introduce the concept of an ideal estimator. This concept is motivated by the idea of an oracle estimator in penalized estimation and selection (Fan and Li (2000)). We use it as the gold standard in our theoretical analysis
and show that the SPIDR estimator is ideal with high probability under
a sparsity and other appropriate conditions.
This implies that the SPIDR estimator is asymptotically normal.
We also illustrate two interesting additional features of SPIDR observed
from our simulation studies: stableness and stickiness. Here by stableness we mean that SPIDR is not sensitive to the change in the penalty parameter within a reasonable range, and by stickiness we mean that the selection
depends on the signal strength of the predictors and  is not severely affected by  the pairwise correlations among the predictors.

Below, we first describe the SPIDR estimator.
We then use a threshold rule for variable selection based on the
SPIDR $z$-statistics and apply the approach for direct FDR control (Storey (2002)) to determine the selection rule.
The details are given in Section \ref{method}, where we also point out that SPIDR naturally leads to confidence intervals for the selected coefficients with a proper confidence statement. In Section \ref{theory} we show that the SPIDR estimator equals an ideal estimator with high probability and describe a stickiness feature of SPIDR.
In Section \ref{NS} we conduct simulation studies to evaluate the finite sample performance of SPIDR and demonstrate its application
on a breast cancer gene expression data set. Section 5 includes some concluding remarks. Proofs of the theoretical results are given in the Appendix.

\section{Method}
\label{method}

\subsection{Semi-penalized estimation}
Let $\bbeta_{-j}=(\beta_k, k\neq j,1\le k \le p )'$
and $X_{-j}=(\bx_k, k\neq j,1\le k\le p)$.
Consider the semi-penalized criteria
\begin{equation}
\label{SP1}
L_j(\bbeta; \lam) = \frac{1}{2n}\|\by-\bx_j\beta_j -X_{-j}\bbeta_{-j}\|^2
+ \sum_{k\neq j} \rho(\beta_k;\lam), 1\le j \le p,
\end{equation}
where $\rho$ is a penalty function with a tuning parameter $\lam \ge 0$.
With these semi-penalized criteria, we concentrate on each coefficient $\beta_j$ one at a time.  The penalization in (\ref{SP1}) is used  to deal with the high-dimensionality of the model. Indeed, the selection of the variables in $X_{-j}$ is to assist with the estimation of $\beta_j$.

We focus on the MCP (Zhang 2010),
\begin{equation}
\label{mcp1}
\rho(t;\lam) = \lam \int_0^{|t|} \Big(1-\frac{x}{\gam\lam}\Big)_+ dx.
\end{equation}
where $\gam$ is a given parameter that controls the concavity of $\rho$.
Here $a_+\equiv a1\{a> 0\}$ is the positive part of $a \in \real$.
The MCP converges to the $\ell_1$ penalty as $\gam \to \infty$ and to the hard threshold penalty as $\gam \to 1$. So the Lasso and hard threshold penalties can be considered two extremes of the MCP with $\gam \to \infty$ and $\gam \to 1$, respectively. A detailed analysis of the MCP is given
in Zhang (2010). We note that other penalized methods such as SCAD and
adaptive Lasso (Zou 2006) can also be  used.

For a fixed $\lam$, let $\hbbeta_{(j)}(\lam)=(\hbeta_j(\lam), \hbbeta_{-j}(\lam))$ be the value that minimizes
the $j$th penalized criterion in (\ref{SP1}), that is,
\begin{equation}
\label{SP2}
\hbbeta_{(j)}(\lam)=(\hbeta_j(\lam), \hbbeta_{-j}(\lam)) = \argmin_{\beta_j, \bbeta_{-j}} L_j(\bbeta;\lam), 1\le j \le p.
\end{equation}
Let $Q_j=I-\bx_j(\bx_j'\bx_j)^{-1}\bx_j'$.
It can be easily verified that
\begin{equation}
\label{SP2a}
\hbbeta_{-j}(\lam) = \argmin_{\bbeta_{-j}}
\frac{1}{2n}\|Q_j(\by -X_{-j}\bbeta_{-j}\|^2
+ \sum_{k\neq j} \rho(\beta_k;\lam),
\end{equation}
and
\begin{equation}
\label{SP2b}
\hbeta_j(\lam) = \argmin_{\beta_j} \|\by-X_{-j}\hbbeta_{-j}-\bx_j\beta_j\|^2 =(\bx_j'\bx_j)^{-1}\bx_j'(\by-X_{-j}\hbbeta_{-j}(\lam)).
\end{equation}
Thus  $\hbeta_j$ is the least squares estimator based on the residuals $\by-X_{^{-j}}\hbbeta_{^{-j}}$ versus $\bx_j$.
Let $\hS_j=\{k: |\hbeta_k(\lam)|>0, k \neq j\}$ be the set of nonzero elements in $\hbbeta_{^{-j}}$. We can write
\begin{equation}
\label{hbeta1}
\hbeta_j(\lam) = (\bx_j'\bx_j)^{-1}\bx_j'(\by-X_{\hS_j}\hbbeta_{\hS_j}(\lam)).
\end{equation}
Here and in the sequel we use the notation $X_A=(\bx_j: j \in A)$ and $\bbeta_A=(\beta_j: j\in A)'$ for any $A \subset \{1, \ldots, p\}$.
Take all the $\hbeta_j(\lam)$'s  as a whole and denote it by $\hbbeta(\lam) = (\hbeta_1(\lam), \ldots, \hbeta_p(\lam))'$.
For simplicity, we refer to $\hbeta(\lam)$ as a SPIDR estimator.
SPIDR estimates one component of $\bbeta$ at a time. This is similar to how spiders make their webs by adding one layer of thread at a time.

In comparison, the fully penalized criterion is
\begin{equation}
\label{FP1}
L(\bb;\lam)
 = \frac{1}{2n}\|\by-\sum_{j=1}^p \bx_j b_j\|^2
+ \sum_{j=1}^p \rho(b_j;\lam).
\end{equation}
For a given $\lam$, the solution to (\ref{FP1}) is $\hbb(\lam)=\argmin_{\bb}L(\bb;\lam)$.
Usually, a $\lam=\hlam$ is chosen based a data-driven procedure such as cross validation. Then $\hbb(\hlam)$ is the penalized estimator of $\bbeta$. Since $\hbb(\hlam)$ can take exact zero value, the set
 $\hS^*=\{j: |\hb_j(\hlam)|>0, 1 \le j\le p\}$ is taken as
an estimator of $S$ based on the fully penalized criterion (\ref{FP1}).

We use a simple example to illustrate the basic properties of the solution paths $\hbbeta_{(j)}(\lam)$ and see how they differ from the fully penalized solution $\hbb(\lam)$.  Consider (\ref{LinMod1})
with $(\beta_1, \ldots, \beta_6)=(3,2,1,-0.5,-1.0,-1.5)$, $\beta_j=0, 7\le j\le p$ and error distribution $N(0, 2.5^2)$.
We set $n=100, p=1000$.
Let $\{z_{ij}, 1\le i\le n, 1\le j\le p\}$ and $\{u_{ij}: 1\le i\le n, j=1,2\}$ be independently generated random numbers from $N(0,1)$.
The predictors are
\begin{align*}
&x_{ij}= z_{ij}+a u_{i1}, j=1,\ldots, 4,  \
x_{ij}=z_{ij}+ a u_{i2}, j=5, \ldots, 8, \\
&x_{ij}=z_{ij}+ u_{i1}, j=9, \ldots, 17, \
x_{ij}=z_{ij}+ u_{i2}, j=18, \ldots, 26, \
x_{ij}=z_{ij}, j=27, \ldots, p.
\end{align*}
We consider two values of $a$, $a=\sqrt{1/3}$ and $a=1$. The strength
of the correlation between the predictors are determined by $a$.
The maximum correlation is $r=a^2/(1+a^2)$. So for $a=\sqrt{1/3}$, $r=0.25$ and for $a=1, r=0.5$.

\begin{figure}[H]
 \centering
 \includegraphics[width=\textwidth]{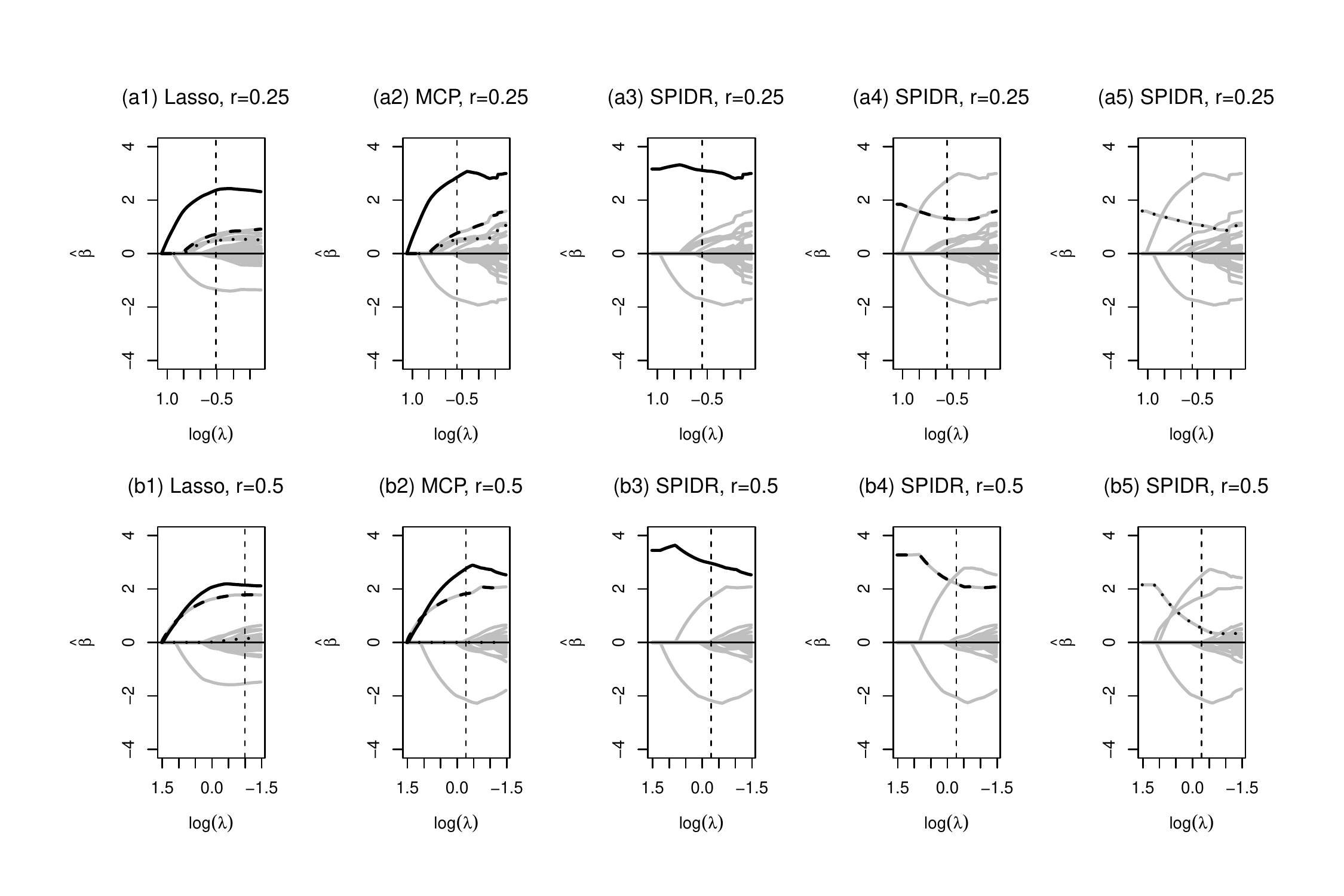}
 \caption{\label{fig1}
 Lasso, MCP and SPIDR solution paths. The results for $r=0.25$ are shown in the top panel (a1)-(a5),
where (a1) and (a2) show the Lasso and
MCP solution paths; (a3)-(a5) show the semi-MCP solution paths of $\hbbeta_{(1)}$, $\hbbeta_{(2)}$ and $\hbbeta_{(3)}$.
The solid, dashed and dotted lines represent the paths of $\hbeta_1$,
$\hbeta_2$ and $\hbeta_3$,
corresponding to $\beta_1=3$, $\beta_2=2$ and $\beta_3=1$,
respectively. The bottom panel (b1)-(b5) in Figure 1 shows the results
for $r=0.5$. The vertical lines are at the value of $\lam$
chosen based on 5-fold cross validation.
}
\end{figure}

The solution paths for $r=0.25$ are shown in the top panel of Figure \ref{fig1}, where (a1) and (a2) show the Lasso and
MCP  paths, respectively;  (a2)-(a5) show the SPIDR solution paths $\hbbeta_{(1)}$, $\hbbeta_{(2)}$ and $\hbbeta_{(3)}$.
The solid, dashed and dotted lines represent $\hbeta_1$,
$\hbeta_2$ and $\hbeta_3$,
corresponding to $\beta_1=3$, $\beta_2=2$ and $\beta_3=1$,
respectively. The bottom panel in Figure 1 shows the results
for $r=0.5$. The vertical lines are at the value of $\lam$
chosen based on 5-fold cross validation.
In (a1), $\log(\hlam)=-0.47$, in (a2)-(a5),  $\log(\hlam)=-0.34$.
In (b1), $\log(\hlam)=-0.99$, in (b2)-(b5),  $\log(\hlam)=-0.27$.

This example illustrates two important features of the SPIDR estimator.
First,  the SPIDR estimator is stable with respect to the change in the penalty parameter.
This intuitively makes sense since $\hbeta_j$ is not subject to penalization. Second, the SPIDR solution paths are less severely impacted by the correlation among predictors. Indeed, it can be seen in Figure 1 (a1) and (b1) as correlation increases from 0.25 to 0.5, it becomes more difficult for Lasso and MCP to correctly select variables with smaller coefficients. But the SPIDR estimator is still able to identify such variables. So the pairwise correlations among the predictors do not have an impact on the SPIDR estimator as big as on the Lasso or MCP.
We refer to this feature of the SPIDR estimator as stickiness.
We will give a formal description of it in Section \ref{theory}.


\subsection{Selection with direct false discovery rate control}
\label{FDR}
In this subsection, we first give a heuristic argument for the
distributional property of $\hbbeta$. We then use this property
to define a selection rule based on directly controlling false discovery rate. We also discuss the confidence intervals of the selected coefficients that can be considered dual to the selection results.

For $A\subset \{1,\ldots, p\}$, denote the projection matrix onto the column space of $X_A$
by $P_A=X_A(X_A'X_A)^{-}X_A'$. Let $Q_{^{\hS_j}} = I - P_{^{\hS_j}}$
and let $\Sigma_{\hS_j}=X'_{\hS_j}X_{\hS_j}/n$.
Suppose the value of the penalty parameter $\lam$ is chosen using
cross validation.
Let $\hbeta_j = \hbeta_j(\lam)$. A useful alternative expression of (\ref{hbeta1}) for $\hbeta_j$ is
\begin{equation}
\label{hbeta2}
\hbeta_j = (\bx_j'Q_{\hS_j}\bx_j)^{-1}\bx_j'[Q_{\hS_j} \by
+ X_{\hS_j}\Sigma_{\hS_j}^{-1}\drho(\hbbeta_{\hS_j};\lam)],
\end{equation}
where $\drho(\hbbeta_{\hS_j};\lam)\equiv (\drho(\hbeta_j;\lam): j \in \hS_j)'$. We verify (\ref{hbeta2}) in the Appendix.

We can write (\ref{hbeta2}) as
\[
\hbeta_j =
(\bx_j'Q_{\hS_j}\bx_j)^{-1}\bx_j'Q_{\hS_j}\by + (\bx_j'Q_{\hS_j}\bx_j)^{-1}\bx_j'X_{\hS_j}
\Sigma_{\hS_j}^{-1}\drho(\hbbeta_{\hS_j};\lam),
\]
where the second term on right hand side represents the bias introduced
by correlation between $\bx_j$ and $X_{\hS_j}$ and penalization.
If this correlation is small, then the bias is negligible.
In general, if the nonzero coefficients are bigger than $\gam\lam$
and the estimator $\hbbeta_{^{\hS_j}}$ is consistent so that
$\hbeta_j \ge \gam\lam$ for all $j \in \hS_j $ with high
probability, then since the derivative of MCP
\(
\drho(t;\lam) = \lam \{1-|t|/(\gam\lam)\}_{+}\sgn(t),
\)
$\drho(\hbbeta_{\hS_j};\lam) =0$ with high probability.
In addition, if the estimator based on (\ref{SP2})
is selection consistent in the sense that $\hS_j$ equals $S_j\equiv \{k: \beta_k\neq 0, k\neq j\}$ with high probability, then
\begin{equation}
\label{App1}
\hbeta_j \approx (\bx_j'Q_{S_j}\bx_j)^{-1}\bx_j'Q_{S_j}\by, 1\le j\le p.
\end{equation}
In Section \ref{theory} we provide sufficient conditions under which the approximations in (\ref{App1})
hold simultaneously for all $1\le j\le p$ with high probability.
Under model (\ref{LinMod1}), $\by=\bx_j\beta_j + X_{S_j}\bbeta_{S_j}+\bveps$, so we have
\[
 \hbeta_j \approx  \beta_j + (\bx_j'Q_{S_j}\bx_j)^{-1}\bx_j'Q_{S_j}\bveps.
\]
It follows that $\hbeta_j$ is consistent and asymptotically normal. Its variance can be consistently estimated by
\begin{equation}
\label{Var1}
\hsigma_j^2 = \hsigma^2 (\bx_j'Q_{^{\hS_j}}\bx_j)^{-1},
\end{equation}
where $\hsigma^2$ is a consistent estimator of $\sigma^2$. We describe an
approach for obtaining such an estimator in Section \ref{NS}.
The covariance between $\hbeta_j$ and $\hbeta_k$  can be
consistently estimated by
\begin{equation}
\label{Cov1}
\widehat{\cov}(\hbeta_j, \hbeta_k)= \hsigma^2 \frac{\bx_j'Q_{^{\hS_j}}Q_{^{\hS_k}}\bx_k}{
(\bx_j'Q_{^{\hS_j}}\bx_j)(\bx_k'Q_{^{\hS_k}} \bx_k)}.
\end{equation}

Thus $\hbeta=(\hbeta_1,\ldots, \hbeta_p)'$ has an asymptotic
multivariate normal distribution with mean $(\beta_1, \ldots, \beta_p)'$
and covariance matrix specified by (\ref{Var1}) and (\ref{Cov1}).
This enables us to formulate the problem of variable selection into the
framework of large scale hypothesis test.

We consider the $z$-statistics $z_j = \hbeta_j/\hsigma_j, 1\le j \le p$.
We can think of variable selection as testing $p$ hypotheses
$H_{0j}: \beta_j=0, 1\le j\le p$.
For a given $t>0$, we reject $H_{0j}$ if $|z_j|>t$, or equivalently,
we select the $j$th variable if $|z_j|>t$. Therefore, the problem
of variable selection becomes that of determining a threshold value
according to a proper control of error.
Let $R(t)= \sum_{j=1}^p 1\{|z_j|>t\}$ be the number of  variables with $|z_j|> t$, and let $V(t)= \sum_{j=1}^p 1\{|z_j|>t, \beta_j=0\}$ be the number of falsely selected variables. We can also write
$V(t) =\sum_{j\in S^c}1\{|z_j|>t\}$, where $S^c$ is the complement of
$S$ in $\{1, \ldots, p\}$.

The false discovery proportion, or the proportion of the null variables among the selected ones for a given $t$ is
\begin{equation}
\label{Fdp1}
\Fdp(t) = \left\{ \begin{array}{cc}
      \frac{V(t)}{R(t)} & \mbox{ if } R(t) >0, \\
      0                 & \mbox{ if } R(t) = 0.
      \end{array}
      \right.
\end{equation}
The FDR is defined to be $Q(t)=\rE(\Fdp(t))$ (Benjamini and Hochberg (1995)). We seek a selection rule $R(t)$ by directly controlling
$Q(t)$. This approach was first proposed in the context of multiple comparisons by Storey (2002).
In theory, we can choose a threshold $\tt_q$ such that $Q(\tt_q)$ equals  a given $0<q<1$.
However, since $Q(t)$ is an unknown population quantity, we need to estimate it in order to determine the threshold value.
We can not directly use
$\Fdp(t)$ as an estimator of $Q(t)$, since $V$ is unobservable.
An approximation to $V(t)$ is by its expectation,
$\rE V(t) \approx 2 |S^c| \Phi(-|t|)$, where $\Phi$ is the
standard normal distribution function.
In sparse models with $|S^c|/p\approx 1$, we further approximate $V(t)$ by $\hV(t)=2p \Phi(-|t|)$.  This results in a first estimate of the FDR
\begin{equation}
\label{Fdr1}
\hQ_0(t) =\left\{ \begin{array}{cc}
      \frac{\hV(t)}{R(t)} & \mbox{ if } R(t) >0, \\
      0                 & \mbox{ if } R(t) = 0.
      \end{array}
      \right.
\end{equation}

For independent test statistics, $\hQ_0$ is a good estimator of
$Q$. However, for correlated statistics, Efron (2007) demonstrated that $\hQ_0$ can give grossly misleading estimate of FDR and proposed an improved estimator. For two-sided tests, this estimator is
\begin{equation}
\label{Fdr2}
\hQ(t) = \hQ_0(t)\left[1+2A\frac{t \phi(t)}{\sqrt{2}\Phi(-t)}\right],
\end{equation}
where $\hQ_0(t)$ is given in (\ref{Fdr1}), $\phi$ is the probability density function of $N(0,1)$. Here $A$ is a dispersion variable accounting for the correlation of the statistics $\hz_j$, which can be estimated based on the their observed values. Methods for estimating
$A$ are given in Efron (2007).

For $0<q < 1$, let $\hatt_q$ be the value satisfying $\hQ(\hatt_q)=q$,
which is an estimator of $\tt_q$.
The set of the indices of the selected variables is
\begin{equation}
\label{SR1}
\hS_q=\{j: |z_j|\ge \hatt_q\}.
\end{equation}
By construction, the FDR of $\hS_q$ is approximately controlled at the level $q$.

\subsection{Confidence intervals of selected coefficients}
The selection rule (\ref{SR1}) directly leads to confidence intervals
for the coefficients of the selected variables.
The $1-q$ level FDR-adjusted confidence intervals of the selected coefficients are
\begin{equation}
\label{CI1}
\hbeta_j \pm \hatt_q\hsigma_j , j \in \hS.
\end{equation}
The interpretation is that the expected
proportion of the these intervals that do not cover
their respective parameters is $q$. Benjamini and Yekutieli (2005)
systematically studied the problem of constructing confidence intervals
for selected parameters and proposed the false coverage-statement rate (FCR) as a measure of interval coverage following
selection. In the present setting, the FCR is exactly the same as
the FDR and the confidence intervals given in (\ref{CI1}) are dual to the selection rule (\ref{SR1}).

\begin{figure}[H]
\centering
\includegraphics[width=\textwidth]{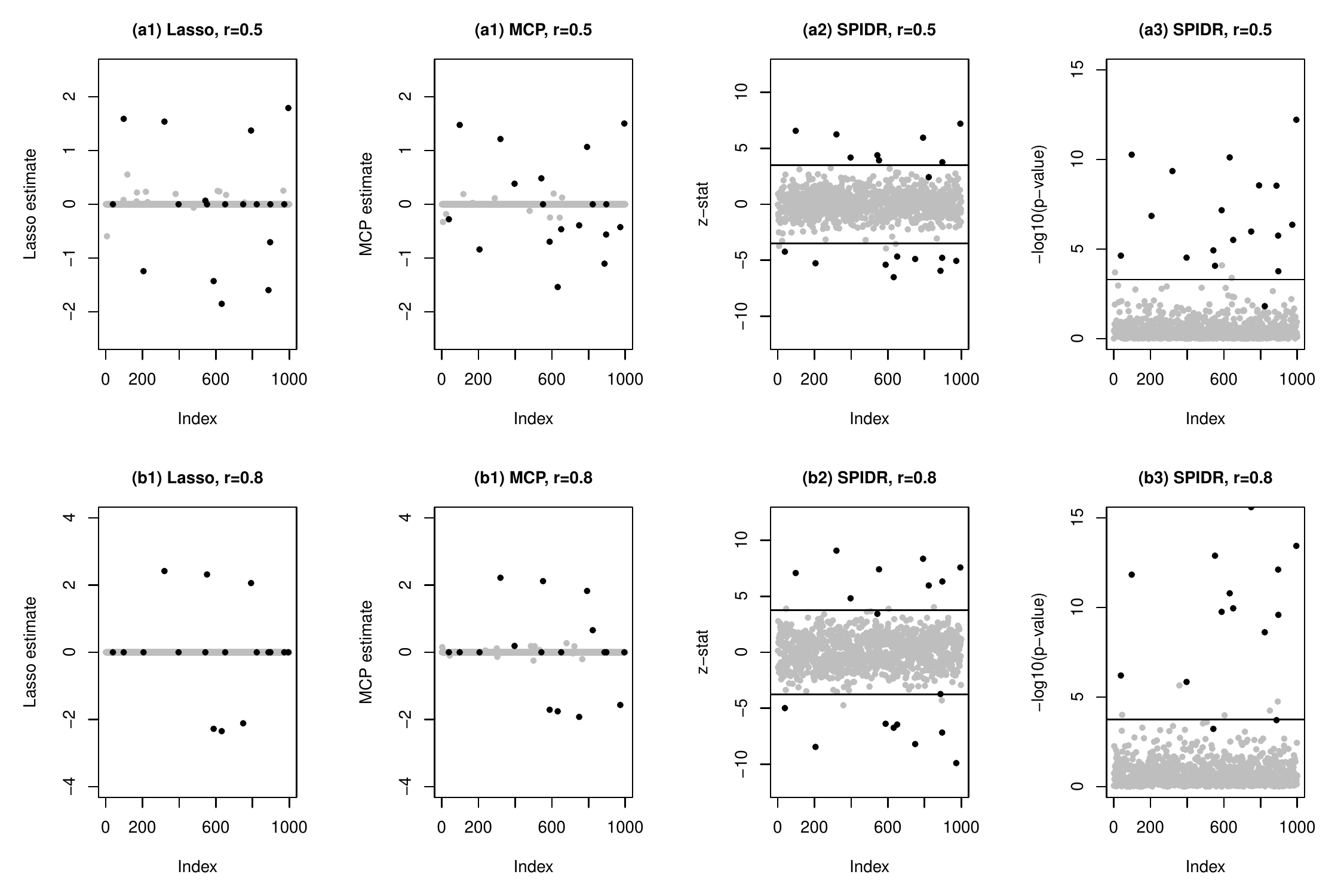}
\caption{\label{fig2} Selection results with $q=0.15$  from the models in Examples 1 and 2.
The top panel (a1)-(a4) shows the results from Example 1 with correlation $r=0.5$.
(a1) and (a2): the Lasso MCP selection results, the black dots represent predictors with nonzero coefficients; (a3): the $z$ statistics based on the SPIDR, the two horizontal lines are drawn at $\pm \hatt_q$.
The bottom panel (b1)-(b4) shows the results from Example 2 with
correlation $r=0.8$.
}
\end{figure}

As an illustration of SPIDR selection and confidence intervals,
Figure \ref{fig2} shows the $z$-statistics and $p$-values based on
simulated data from the two models described in Examples 1 and 2 in Section \ref{NS}. For comparison, we also include the selection results from the Lasso and MCP. In these two examples, there are $18$ predictors with nonzero coefficients among a total of $p=1000$ variables. Here the indices of the nonzero coefficients are randomly selected from $1$ to $p$.
The top panel in Figure \ref{fig2} shows the results from a model with the largest pairwise correlation $r=0.5$, where (a1) and (a2) show the Lasso and MCP selection results, the black dots represent predictors with nonzero coefficients; (a3) shows the SPIDR $z$-statistics, the two horizontal lines are drawn at the threshold values $\pm \hatt_q$ with $\hatt_q=3.48$ and $q=0.15$; and (a4) shows the negative $\log_{10}$ of the $p$ values based on the $z$ statistics, the horizontal line is drawn at $-\log_{10}(2\Phi(-\hatt_q))=3.30$.
Plots (b1)-(b4) in the bottom panel show the results from Example 2 with
$\hatt_q=3.75$ in (b3),  $-\log_{10}(2\Phi(-\hatt_q))=3.75$ in (b4) and  the largest pairwise correlation $r=0.8$.

By examining Figure \ref{fig2}, we see that SPIDR has better selection performance than Lasso and MCP for these two data sets. For $r=0.5$, it has a smaller FDR and misses fewer non-null predictors. For $r=0.8$, Lasso has zero FDR, but it misses 12 of the 18 non-null predictors. MCP has a higher FDR than SPIDR and misses 9 non-null predictors. It is interesting to note that the performance of SPIDR  remains essentially unchanged as
correlation increases from 0.5 to 0.8. This again illustrates the stickiness feature of SPIDR mentioned earlier. Of course, these observations are based on a single data set. In the simulation studies reported in Section \ref{NS}, they remain true based on replicated simulations.

The difficulty that Lasso has in the presence of high pairwise correlations had been pointed out by Zou and Hastie (2006). This is one of the main motivations for them to introduce the elastic net, which
has a grouping effect by selecting or dropping strongly correlated predictors together.
As described in Section \ref{theory} below, the stickiness feature of SPIDR is different from the grouping effect of the elastic net. It depends on the signal strengths of the variables and residual correlations between predictors, but not the usual pairwise correlations.

\medskip
\begin{figure}[H]
\centering
\includegraphics[width=\textwidth]{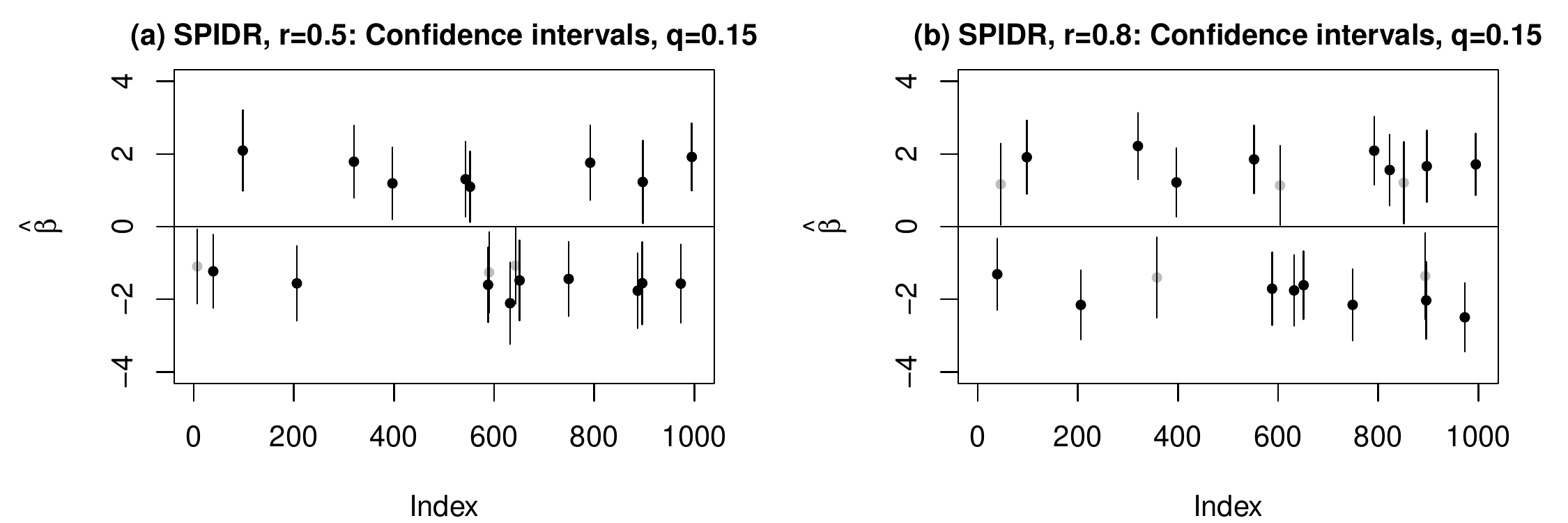}
\caption{\label{fig3} The confidence intervals of the selected coefficients with the FDR level $q=0.15$. The gray dots indicate false
coverage. (a) Confidence intervals for the selected coefficients when $r=0.5$; (b) Confidence intervals for the selected coefficients when $r=0.8$.}
\end{figure}

Figure \ref{fig3} shows the $1-q$ level FDR-adjusted confidence intervals of the selected coefficients with $q=0.15$.
The top plot (a) shows the confidence intervals for the selected coefficients in the model with $r=0.5$ in Example 1 and
the bottom plot (b) shows the results from the model with strongly correlated predictors in Example 2 in Section \ref{NS}.
 The gray dots indicate falsely selected variables. So their corresponding confidence intervals do not cover the true parameters, which are zero. In (a) or (b), approximately 15\% of the intervals will not cover their
 corresponding parameter values.

\section{Theoretical properties}
\label{theory}
In this section, we study the theoretical properties of the SPIDR estimator. We introduce the concept of an \textit{ideal} estimator.
We provide sufficient conditions under which the SPIDR estimator equals
the ideal estimator with high probability.  Consequently
the SPIDR estimator is asymptotically  normal with mean $\bbeta$ and covariance matrix specified by (\ref{Var1}) and (\ref{Cov1}). We also discuss the notion of stickiness we mentioned earlier.

\subsection{Idealness property}

Let $S_j=\{k:\beta_k\neq 0,k\neq j, 1\le k \le p\}$ and let $S_j^c$ be the complement of $S_j$ in $\{1, \ldots, p\}$.
We define the \textit{ideal} estimator by
\begin{equation}
\label{OP1}
(\tbeta_j, \tbbeta_{-j}) = \argmin_{\beta_j, \bbeta_{-j}}
\{\|\by-\bx_j\beta_j-X_{-j}\bbeta_{-j}\|^2: \bbeta_{S_j^c}=0\},
1\le j\le p.
\end{equation}
In particular, $\tbeta_j$ is an ideal estimator of $\beta_j$.
We note that (\ref{OP1}) is a counterpart of (\ref{SP1}) without
penalization assuming that the support of $\bbeta_{-j}$ is known.
It can be verified that an explicit expression of the ideal estimator $\tbeta_j$
is
\begin{equation}
\label{OP2}
\tbeta_j=\beta_j + (\bx_j'Q_{S_j}\bx_j)^{-1}\bx_j'Q_{S_j}\bveps, \quad
(j=1,\ldots, p).
\end{equation}
This expression of $\tbeta_j$ is parallel to (\ref{hbeta2}).
By (\ref{OP2}),


By (\ref{OP2}), $(\tbeta_1, \ldots, \tbeta_p)$ has a multivariate normal distribution with mean vector
$\bbeta$ and
\[
\Var(\tbeta_j) = \sigma^2 (\bx_j'Q_{{S_j}}\bx_j)^{-1} \mbox{ and }
\Cov(\tbeta_j, \tbeta_k)= \sigma^2 \frac{\bx_j'Q_{{S_j}}Q_{{S_k}}\bx_k}{
(\bx_j'Q_{{S_j}}\bx_j)(\bx_k'Q_{{S_k}} \bx_k)}.
\]

We first state a result when the penalized criterion (\ref{FP1}) is
convex. This necessarily requires $p < n$, but allows  $p \to \infty$ as $n \to \infty$.
Let $c_{\min}=\min\{c_j: 1\le j\le p\}$, where $c_j$ is the smallest eigenvalue of $X_{-j}'Q_jX_{-j}/n$.
Let $w^o=\max\{w_{jk}^o: k\in S_j, 1\le j\le p\}$, where
$(w_{jk}^o, k \in S_j)$ are the diagonal elements of
$(X_{S_j}'Q_j X_{S_j}/n)^{-1}$. Denote the smallest nonzero coefficient by  $\beta_*=\min\{|\beta_j^o|: \beta_j^o\neq 0, 1\le j\le p\}$.
Denote the cardinality of $S$ by $|S|$.

\begin{theorem}
\label{Thm1}
Suppose that  $\veps_1, \ldots, \veps_n$ are independent and identically distributed  as $N(0, \sigma^2)$. Also, suppose that (a) $\gam>1/c_{\min} $; (b) for a small $\eps >0$, $\beta_*> \gamma \lam+\sig\sqrt{(2/n)w^o\log (p|S|/\eps)}$; and (c) $\lam \ge \sigma \sqrt{4\log p}  \max_{j\le p}\|\bx_j\|/n$.
Then,
\[
\rP\{\cup_{j=1}^p(\hS_j\neq S_j)\} \le 3 \eps \mbox{ and }
\rP\{\cup_{j=1}^p(\hbeta_j(\lam)\neq \tbeta_j)\}
\le 3 \eps.
\]
\end{theorem}
This theorem shows that in the convex case, the SPIDR estimator is
asymptotically ideal, meaning that it equals the ideal estimator with high probability. As a consequence, it is asymptotically normal.
The conditions are mild. The normality assumption on the errors is
mainly used for bounding the tail probabilities of the error distribution. This assumption can be relaxed. Condition (a) guarantees  that the SPIDR criteria in (\ref{SP1}) are strictly convex to ensure unique solution.
Condition (b) requires that the nonzero coefficients not
be too small so that it is possible to separate them from zero in the
presence of random noise. Condition (c) requires the penalty to be proportionally greater than the noise level to prevent false
selection of null variables. For standardized predictors with $\|\bx_j\|^2=n$, this condition simplifies to
$\lam \ge \sigma \sqrt{(4/n)\log p}$. Conditions (b) and (c) are related, a bigger $\lam$ requires a bigger $\beta^*$.

We now consider the high-dimensional cases where $p \gg n$ and the
criteria (\ref{SP1}) are nonconvex. We require the sparse Riesz condition (SRC, Zhang and Huang (2008)) on the the matrices $Q_jX$.
Specifically, we assume there exist constants $0<c_*\le c^*<\infty$
and integer $d^*\ge |S|(K_*+1)$ with $K_*=c^*/c_* -1/2$ such that
\begin{equation}
\label{SRC1}
0 < c_*\le \|Q_j X_{A_j}\bu\|^2/n \le c^*< \infty, \|\bu\|_2=1,
\end{equation}
for every $A_j \subset \{1,\ldots, p\}\setminus\{j\} \mbox{ with } |A_j\cup S_j| \le d^*$,
for all $1\le j \le p$.

\begin{theorem}
\label{Thm2}
Suppose that  $\veps_1, \ldots, \veps_n$ are independent and identically distributed  as $N(0, \sigma^2)$. Also, suppose that (a) the SRC (\ref{SRC1}) holds with $\gamma \ge c_*^{-1}\sqrt{4+c_*/c^*}$;
(b) for a small $\eps >0$, $\beta_* \ge \gam 2\sqrt{c^*}\lam+\sig\sqrt{(2/n)w^o\log(p|S|/\eps)}$;
(c) $\lam \ge \sigma \sqrt{(4\log(p/\eps)} \max_{j\le p}\|\bx_j\|/n$.
Then
\[
\rP\{\cup_{j=1}^p(\hS_j(\hlam)\neq S_j)\} \le 3\eps,
\ \mbox{ and } \
\rP\{\cup_{j=1}^p(\hbeta_j(\hlam)\neq \tbeta_j)\}
\le 3\eps.
\]
Therefore, $\rP\{\cup_{j=1}^p(\hS_j(\hlam)\neq S_j)\} \to 0$
and $\rP\{\cup_{j=1}^p(\hbeta_j(\hlam)\neq \tbeta_j)\} \to 0$ as
$\eps \to 0$.
\end{theorem}

The SRC (\ref{SRC1}) ensures that the model is identifiable in a lower-dimensional space that contains the underlying model. When $p > n$, the smallest eigenvalue of $X_j'Q_jX_j/n$ is always zero. But the requirement $c_*> 0$  only concerns $d^*\times d^*$ diagonal submatrices of $X_j'Q_jX_j/n$.
By examining the conditions (b) and (c), for standardized predictors with $\|\bx_j\|=\sqrt{n}$, we can have
$\log(p|S|/\eps)=o(n)$ or $p=\eps \exp(o(n))/|S|$. Thus for sparse models
with $|S|$ small relative to $n$, Theorem \ref{Thm2} shows that the asymptotic idealness property of the SPIDR estimators continues to hold in high-dimensional settings under the SRC and other suitable conditions.

Theorems \ref{Thm1} and \ref{Thm2} are stated for fixed predictors.
For random predictors, the conditions involving the predictors such as
the SRC (\ref{SRC1}) need to hold with high probability.

\subsection{Stickiness}

Stickiness refers to a ``robustness'' property of a selection rule with respect to pairwise correlations among predictors. Specifically,
a selection rule is sticky if it is capable of catching a variable with
a relatively big coefficient, even if it is highly correlated with
some other predictors.

In SPIDR, selection is based on the $z$-statistics $z_j, 1\le j\le p$.  Variables with similar $z$-statistic values will be selected or dropped together. So we examine the difference between $z_j$ and $z_k$ for $j\neq k$. Based on the asymptotic idealness property of SPIDR stated in
Theorem \ref{Thm1} or Theorem \ref{Thm2}, we can look at the ideal
estimator from a large sample standpoint.

We first consider the notion of signal strength for measuring the
importance of a predictor.
Let $m_j=(\bx_j'Q_{S_j}\bx_j)^{1/2}$.
The ideal estimator of $\beta_j$ can be written as
$\tbeta_j=m_j^{-1}\bx_j'Q_{S_j}\by$.
The corresponding $z$-score is
\(\tz_j = m_j (\tbeta_j/\sigma).\)
We define the signal strength of the $j$th predictor by
\[
\psi_j = \rE \tz_j = m_j (\beta_j/\sigma).
\]
The interpretation of $\psi_j$ is clear, it depends on the ratio of the $j$th coefficient over the error standard deviation and the length of $Q_{S_j}\bx_j=\bx_j-P_{S_j}\bx_j$, the vector of residuals of $\bx_j$ regressing on the variables in $S_j$. We refer to $\beta_j/\sigma$ as the
base signal and $m_j$ as the signal multiplier.

In the extreme case where the signal multiplier $m_j$ is zero,
that is, $\bx_j$ is perfectly correlated with the variables in $S_j$, the signal strength of $\bx_j$ is zero, no matter how large the base signal is. On the other hand, for a variable with a small to moderate base signal $\beta_j/\sigma$, its signal strength can still be large if its signal multiplier is large.

With the definition of signal strength, we can now propose a measure
of stickiness. Specifically, we measure stickiness by the root mean squared difference
\(
\ts_{jk}\equiv \{\rE(\tz_j-\tz_k)^2\}^{1/2}.
 \)
It can be easily verified that
\begin{equation}
\label{sticky1}
\ts_{jk}^2= (\psi_j-\psi_k)^2+2(1-\Cov(\tz_j,\tz_k)),
\end{equation}
where
\[
\Cov(\tz_j,\tz_k)=\Cor(\tbeta_j, \tbeta_k)= \frac{\bx_j'Q_{{S_j}}Q_{{S_k}}\bx_k}{
(\bx_j'Q_{{S_j}}\bx_j)^{1/2}(\bx_k'Q_{{S_k}} \bx_k)^{1/2}}.
\]
So stickiness is determined by the difference in signal strengthes and the predictor residual correlation between $Q_{S_j}\bx_j$ and $Q_{S_k}\bx_k$. It is not related to the usual pairwise correlations.
Signal strength is a main factor in determining SPIDR selection. The pairwise correlations among predictors do not have an impact as big as in penalized selection.
By considering stickiness, we identified two key quantities that affect SPIDR selection: the signal strength and pairwise predictor residual correlation.

\section{Numerical studies}
\label{NS}
\subsection{Implementation}
To implement the proposed method, we need to determine the penalty
parameter $\lam$ and estimate the error variance $\sigma^2$. The former
is needed for estimating the regression coefficients and the latter
is required for computing the $z$-statistics based on the estimated
regression coefficients.

We employ $5$-fold cross validation for choosing
$\lam=\hlam$ based on the fully penalized criterion in (\ref{FP1})
using the MCP (\ref{mcp1}) with $\gam=6$.
This requires computing the solution path $\hbb(\lam)=\argmin_{\bb}L(\bb;\lam)$ for $\lam$ in a properly specified interval. The R package \textsl{ncvreg} is used
in the computation. This package implements a coordinate descent algorithm for penalized methods including the Lasso and MCP, and is available at \url{cran.r-project.org/web/packages/ncvreg}
(Breheny and Huang (2009)). This $\hlam$ is then used in calculating $\hbeta_j = \hbeta_j(\hlam), 1\le j\le p$ in (\ref{SP1}). In this way,
it is only necessary to calculate $\hbeta_j$ at $\hlam$.
Conceptually, it is possible
to choose a different $\lam$ for each $\hbeta_j$. However, this will substantially increase the computational cost, since it will involve
calculating the whole solution path for each of the $p$ minimization
problems in (\ref{SP1}). Also,  since $\hbeta_j$ is not very sensitive
to $\lam$, choosing a $\hlam$ based on (\ref{FP1}) is reasonable.

For estimating $\sigma^2$, we use the following procedure.
Let $\hbb(\hlam)$ be the MCP estimator with $\hlam$ determined based
on $5$-fold cross validation.
Let $\hS$ be the set of the predictors with nonzero coefficients in
$\hbb$. We randomly partition the dataset into two subsets $D_1$ and $D_2$ with equal sample sizes $n_1=n_2=n/2$. We use the first part to fit a model with variables in $\hS$ and calculate the least squares estimate
\[
\hbb^{(1)} = \argmin_{\bbeta} \sum_{i\in D_1}(y_i - \sum_{j\in \hS}x_{ij}b_j)^2.
\]
Let
\begin{equation}
\label{hsigma1}
\hsigma^2=\frac{1}{n_2+|\hS|} \sum_{i \in D_2}(y_i-
\sum_{j\in \hS}x_{ij}\hb^{(1)}_{j})^2.
\end{equation}
We show in the Appendix that this is a consistent estimator of $\sigma^2$. To smooth out the variations of the random partition,
we repeat this process 10 times and take
the average of the resulting $\hsigma^2$'s as the estimate of $\sigma^2$.

This procedure bears some resemblance to the cross-refitted method for
variance estimation in Fan et al. (2012). But there are also important
differences. Here we use the full dataset to select variables and then use a properly scaled prediction error for variance estimation. One reason for using the full dataset as opposed to using a subset is to achieve better selection results. Another reason is to take advantage of the
fact that in choosing the penalty parameter $\hlam$ based on (\ref{FP1}) for calculating the SPIDR estimators based on (\ref{SP1}), we have already computed the full penalized estimator. Thus the procedure described above does not incur any significant extra computational burden.

We have also looked at the methods based on ordinary least squares with the variables selected using the MCP criterion (\ref{FP1}) and data partition. The estimator given in (\ref{hsigma1}) is competitive, and in general, it tends to give more accurate estimates. Variance estimation is an important problem in high-dimensional regression. We refer to Fan et al. (2012) and Sun and Zhang (2012) for the discussions on this problem and other approaches.

\subsection{Simulation studies}
We focus on the selection results of the SPIDR method in three models described below. Specifically, we look at the empirical FDR and FMR
(false miss rate). For a given threshold value $t>0$, let
$U(t)=\sum_{j\in S}1\{|z_j|>t\}$ be the number of selected variables in
$S$.
The false miss proportion is defined to be
\[
\Fmp(t)=\frac{|S|- U(t)}{|S|}.
\]
Then the FMR at $t$ is $\rE (\Fmp(t))$.
As a comparison, we also look at the empirical FDR and FMR of the selection results based on the Lasso and MCP.

\noindent\textbf{Example 1.}
We consider model (\ref{LinMod1}) with
$p=1000$. The errors are independent and identically distributed as
$N(0, \sigma^2)$ with $\sigma=3$.
The first $q=18$ coefficients are nonzero with values
\[
(\beta_1, \ldots, \beta_{18}) = (1,1,1,.8,.8,.8,.6,.6,.6,
-.6,-.6,-.6,-.8,-.8,-.8,-1,-1,-1).
\]
The sample size $n=q^2/2=162$. The remaining coefficients are zero.
The predictors are generated as follows.
Let $\{z_{ij}, 1\le i\le n, 1\le j\le p\}$ and $\{u_{ij}: 1\le i\le n, 1\le j \le 2\}$ be independently generated random numbers from $N(0,1)$.
Let $A_1=\{1, \ldots, 9\}$ and $A_2=\{10, \ldots, 18\}$ be the sets of
predictors with nonzero coefficients. Let $A_3, A_4$ and $A_5$ be different sets of 50 indices randomly chosen from $\{19, \ldots, p\}$.
\begin{align*}
&x_{ij}= z_{ij}+a_1 u_{i1}, j\in A_1, \
x_{ij}=z_{ij} + a_1 u_{i2}, j\in A_2, \\
&x_{ij}=z_{ij}+a_2 u_{i1}, j\in A_3, \
x_{ij}=z_{ij}+a_2 u_{i2}, j\in A_4,  \\
&x_{ij}=z_{ij}+a_3u_{i1}-a_3u_{i2}, j\in A_5,
x_{ij}=z_{ij}, j\not\in \cup_{k=1}^5 A_k,
\end{align*}
where
$a_1=1$, $a_2=0.5$ and $a_3=0.1$. In this model, there is correlation
among predictors with nonzero coefficients as well as between such
predictors and predictors with zero coefficients. For example, the correlation of the predictors in $A_1$ is $r_{11}=a_1^2/(1+a_1^2)=0.5$ and the correlation between the predictors in $A_1$ and $A_3$ is
$r_{13}=a_1a_2/(\sqrt{1+a_1^2}\sqrt{1+a_2^2})=0.32$.


\medskip\noindent\textbf{Example 2.}
The generating model is the same as that in Example 1, except
$a_1=2$. Now there is stronger correlation among the predictors. For example, the correlation between the predictors in $A_1$ is $r_{11}=0.8$ and the correlation between the predictors in $A_1$ and $A_3$ is
$r_{13}=0.40$.

\medskip\noindent\textbf{Example 3.} The generating model is the same as that in Example 1, except now the predictors are generated from
a multivariate normal distribution $N(0, \Sigma)$, where the $(j,k)$th
element of the covariance matrix $\Sigma$ is $\sigma_{jk}= 0.5^{|j-k|}$,
$1\le j, k\le p$.

\begin{figure}[H]
\centering
\includegraphics[width=\textwidth]{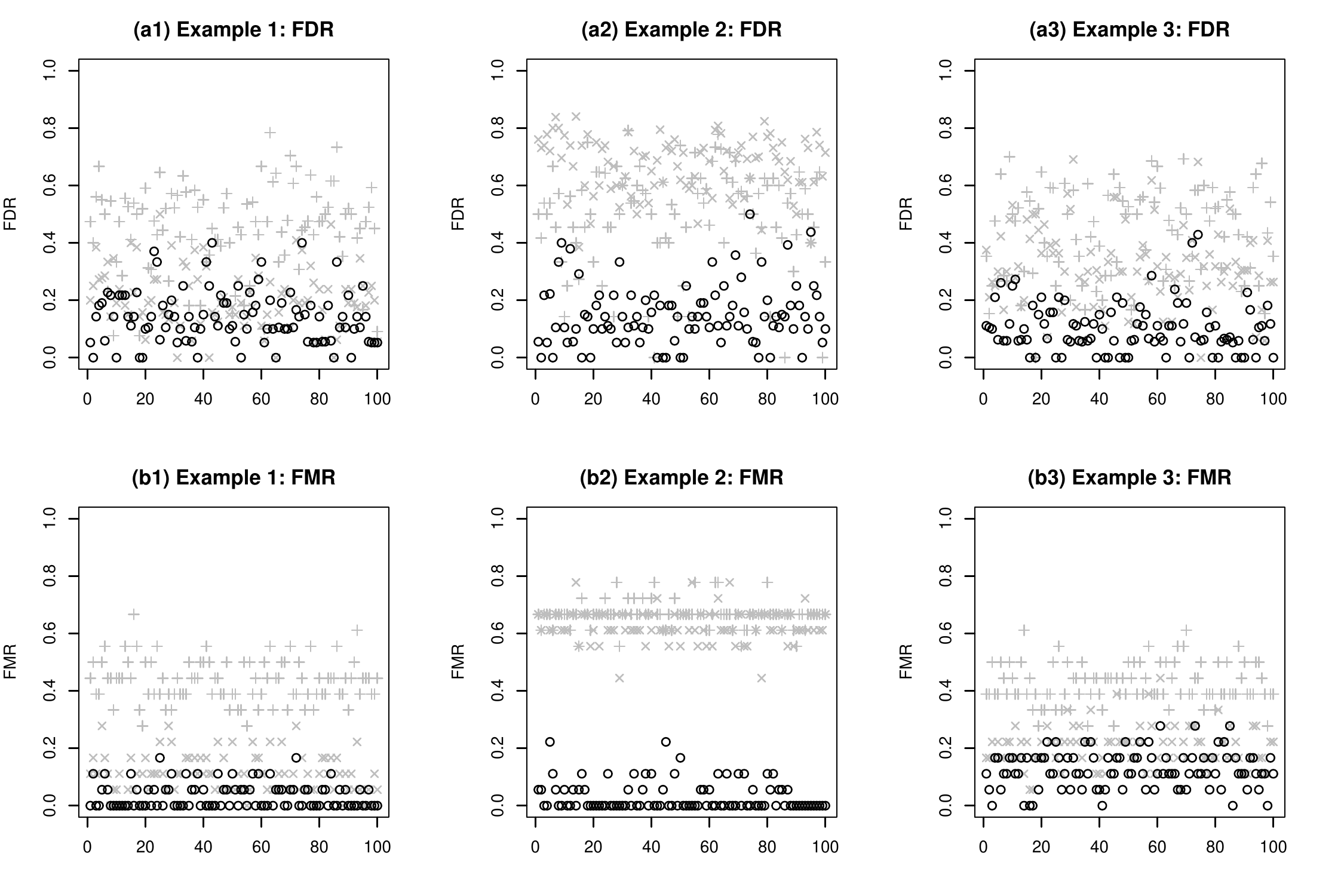}
\caption{\label{fig4} Top panel: False discovery rates from 100 replications for (a1) Example 1, (a2) Example 2 and (a3) Example 3.
Bottom panel: False missing rates from 100 replications for (b1) Example 1, (b2) Example 2 and (b3) Example 3. The results for Lasso, MCP and SPIDR are represented by the plus  ``+'', cross ``x'' and  open circle ``$\circ$''  signs, respectively.}
\end{figure}

Figure \ref{fig4} shows the empirical FDR's and FMR's from 100 replications. For the SPIDR, the nominal FDR  is set at $q=0.15$.
The top panel in Figure \ref{fig4} shows the empirical false discovery rates for (a1) Example 1, (a2) Example 2 and (a3) Example 3, and
the plots (b1)-(b3) in the bottom panel show the empirical false miss rates for these studies. Since it is difficult to assess the absolute performance of the SPIDR, we also include the selection results from the Lasso and  MCP for comparison.  The Lasso and MCP
results are obtained at the penalty parameter value determined by 5-fold cross validation.
In the plots, the results for Lasso, MCP and SPIDR are represented by the plus  ``+'', cross ``x'' and  open circle ``$\circ$''  signs, respectively.

Numerical summaries of Figure \ref{fig4} are given in Table \ref{tab1}.
As can be seen in the plots, there is a fair amount of variations in the false discovery rates. However, the average false discovery rate for SPIDR are close to the nominal level, as shown in Table \ref{tab1}.
Overall, the SPIDR has smaller FDR and FMR than the Lasso and MCP in the
three examples considered. In particular, in Example 2, where the correlation is high, the SPIDR has considerably smaller FDR and FMR than the Lasso and MCP.

\begin{figure}[H]
\centering
\includegraphics[width=\textwidth]{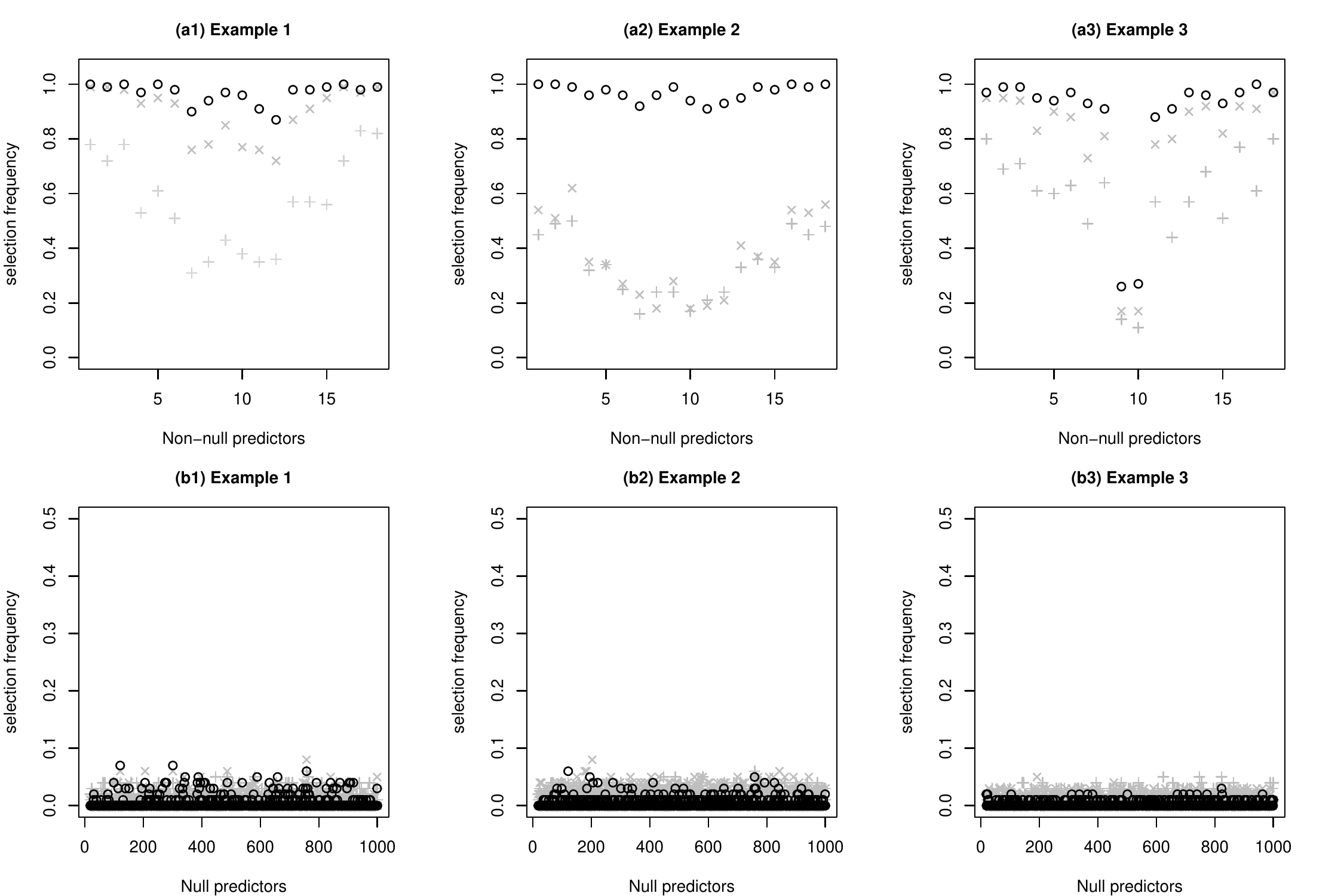}
\caption{\label{fig5} Top panel: percentages of selected non-null variables based on 100 replications for (a1) Example 1, (a2) Example 2 and (a3) Example 3. Bottom panel: percentages of selected non-null variables based on 100 replications for for (b1) Example 1, (b2) Example 2, and (b3) Example 3. The results for Lasso, MCP and SPIDR are represented by the plus  ``+'', cross ``x'' and  open circle ``$\circ$''  signs, respectively.}
\end{figure}

Figure \ref{fig5} shows the percentages of the variables being selected calculated based on 100 replications. The plotting legends are the same as those in Figure \ref{fig4}. The top panel in Figure \ref{fig3} shows the percentages of correct selection (PCS), that is, the  non-null predictors being selected  for (a1) Example 1, (a2) Example 2 and (a3) Example 3. The bottom panel shows the percentages of false selection (PFS), that is,  the null predictors
being selected  for (b1) Example 1, (b2) Example 2 and (b3) Example 3.

\begin{center}
\begin{table}[H]
\caption{\label{tab1} Simulation study.  NVS, number of variables selected; FDR, false discovery rate; FMR, false miss rate,
averaged over 100 replications with standard deviations in parentheses,  for Examples 1 to 3.}

\medskip
\begin{center}
\begin{tabular}{lccc} \hline\hline
Method  &   NVS & FDR & FMR  \\ \hline
& \multicolumn{3}{c}{Example 1}  \\
SPIDR & 20.52 (2.78) & 0.14 (0.09) & 0.03 (0.04)  \\
MCP & 20.66 (2.83) & 0.21 (0.10) & 0.11 (0.07)  \\
Lasso & 19.97 (5.92) & 0.45 (0.15) & 0.43 (0.07)  \\\hline
                    & \multicolumn{3}{c}{Example 2} \\
SPIDR & 20.90 (3.54) & 0.15 (0.11) & 0.03 (0.05) \\
MCP & 21.53 (5.76) & 0.67 (0.10)& 0.63 (0.06) \\
Lasso & 13.22 (4.08) & 0.50 (0.17) & 0.66 (0.04)  \\ \hline
& \multicolumn{3}{c}{Example 3} \\
SPIDR & 17.75 (2.44) & 0.10 (0.08) & 0.12 (0.06) \\
MCP & 22.05 (5.42) & 0.32 (0.14)& 0.20 (0.07) \\
Lasso & 19.63 (6.19) & 0.43 (0.16) & 0.42 (0.07)  \\ \hline \hline

\end{tabular}
\end{center}
\end{table}
\end{center}

In Example 1, the SPIDR has slightly higher PCS and slightly lower
PFS than the MCP. Both SPIDR and MCP perform better than Lasso in terms
of PCS. In Example 2, the SPIDR has considerably higher PCS and
lower PFS than the Lasso and MCP. In Example 3, the SPIDR has higher PCS
and lower PFS than the Lasso and MCP, although for two predictors
with smaller coefficients, all the methods have relatively low PCS.

In summary, the SPIDR has good performance in the examples considered here. It can achieve the nominal FDR control on average and tends to have smaller FMR than the Lasso and MCP.  Especially,  for the model in Example 2, which is a difficult case for the Lasso and MCP because of the high correlations among the predictors, the SPIDR still performs
reasonably well. This demonstrates its stickiness.

\subsection{Data example}
We use the breast cancer data from The Cancer Genome Atlas (2012)
project to illustrate the proposed method.
In this dataset, tumour samples were assayed on several platforms. Here we focus on the gene expression data obtained using Agilent mRNA expression microarrays. In this dataset, expression measurements of 17814 genes, including BRCA1, from 519 patients are available at \url{http://cancergenome.nih.gov/}. BRCA1 is the first gene identified that increases the risk of early onset breast cancer.
Because BRCA1 is likely to interact with many other genes, including tumor suppressors and regulators of the cell division cycle, it is of interest to find genes with expression levels related to that of BRCA1. These genes may be functionally related to BRCA1 and are useful candidates
for further studies.

We only include genes with sufficient expression levels and variations across the subjects in the analysis. So we first do an initial screen
according to the following requirements:
(a) the coefficient of variation is greater than 1;
(b) the standard deviation is greater than 0.6;
(c) the marginal correlation coefficient with BRCA1 is greater than 0.1.
A total of 1685 genes passed these screening steps. These are the genes included in the model.

\begin{figure}[H]
 \centering
 \includegraphics[width=\textwidth]{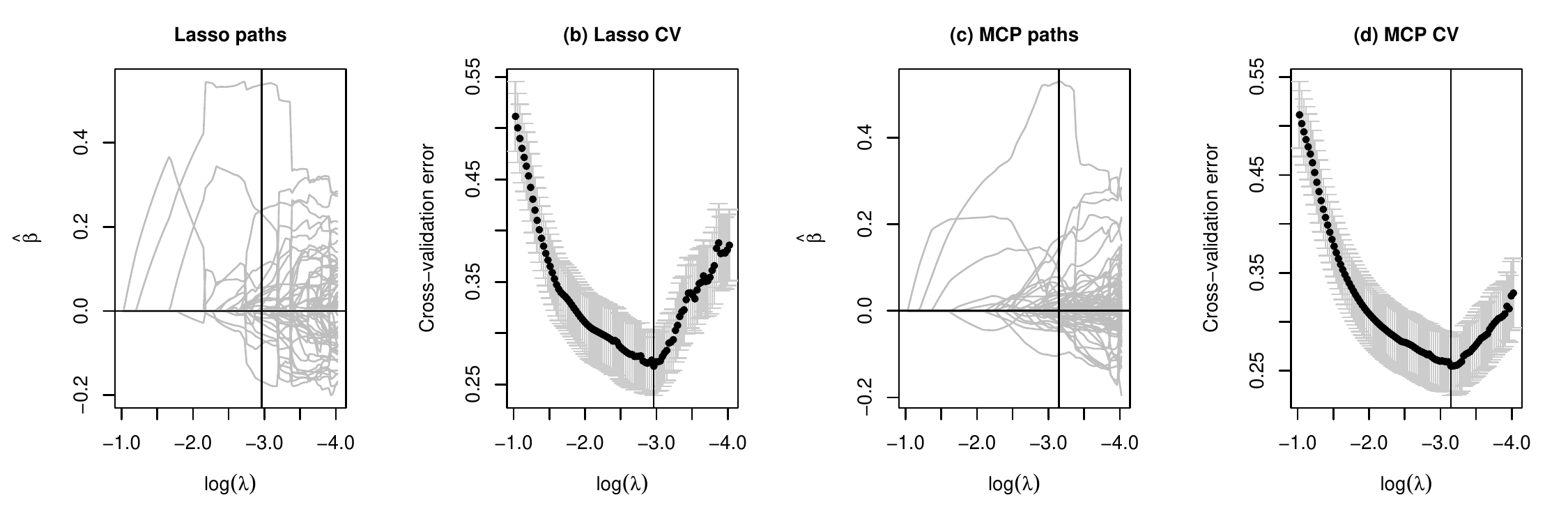}
 \caption{\label{bc-paths1}
 Breast cancer data.  (a) Lasso solution
 paths; (b) Lasso cross validation results; (c) MCP solution paths;
 (d) MCP cross validation results.}
\end{figure}

We start by looking at the Lasso and MCP solution paths together with 5-fold cross validation results, which are shown in Figure \ref{bc-paths1}. The vertical lines are drawn at the values $\hlam$ of the penalty parameter that achieve the smallest cross validation errors for Lasso and MCP, respectively.   For the Lasso, $\log(\hlam) =-2.96 $, for the MCP, $\log(\hlam)=-3.15$. The gray lines in Figure \ref{bc-paths1} (b) and (d) represent the standard deviations of the cross validation errors
calculated based on 5-fold calculations. These plots show that for either
Lasso or MCP, there is a unique point on the solution path that minimizes the cross validation error, which leads to a well-defined model.

\begin{figure}[H]
 \centering
 \includegraphics[width=\textwidth]{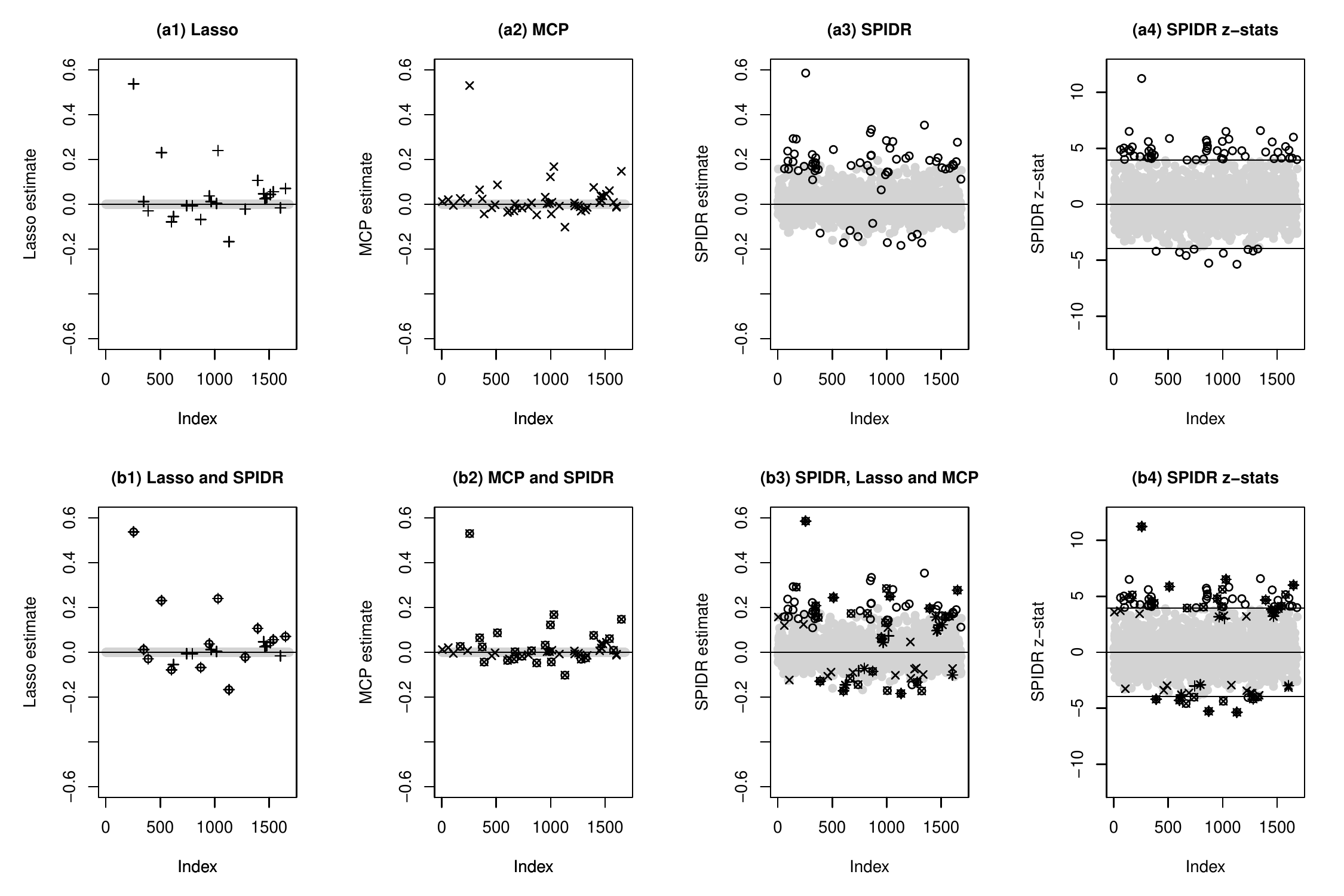}
 \caption{\label{bc-analysis1}
 Breast cancer data. Lasso, MCP and SPIDR are represented by plus ``+'',
 cross ``x'', and circle ``$\circ$", respectively.
 Top panel, (a1): Lasso estimates; (a2) MCP estimates;
 (a3) SPIDR estimates;
 (a4) SPIDR $z$-statistics. Bottom panel, (b1): Lasso and SPIDR overlap;
 (b2) MCP and SPIDR overlap; (b3) SPIDR estimates with Lasso and MCP selection results indicated; (b4) SPIDR $z$-statistics with Lasso and MCP selection
 results indicated.}
\end{figure}

The Lasso and MCP estimates at the cross-validated $\hlam$ are shown
in Figure \ref{bc-analysis1} (a1) and (a2), the plus ``+'' and cross ``x'' signs represent genes selected by Lasso (24 genes) and MCP (48 genes).
Figure \ref{bc-analysis1} (a3) shows the SPIDR estimates, the circles ``$\circ$'' represent the selected genes (63 genes) with $q=0.10$. The SPIDR z-statistics are shown in  (a4), the cut-off values for selection corresponding to FDR level $q=0.10$ are $\hatt_{0.10}=\pm 3.95$, which are indicated by two horizontal lines.
Figure \ref{bc-analysis1} (b1)-(b4) are parallel to (a1)-(a4), but now
the overlaps between the methods are indicated.
Figure \ref{bc-analysis1} (b1) shows the overlap between the Lasso and SPIDR, the circles represent the genes that are also selected by
SPIDR. Similarly,  (b2) shows the overlap between the MCP and SPIDR. In (b3) and (b4), all the selected genes based on the three methods are indicated. As can be seen in (b4), genes with relatively large estimated coefficients based on Lasso or MCP are also selected by SPIDR, whereas those with small estimated coefficients tend to be deemed nonsignificant by SPIDR. There are large overlaps between the three methods. For example, 13 genes are selected by both Lasso and SPIDR,  these same 13 genes
are selected by all the three methods, and there are 24 genes selected by both MCP and SPIDR. One of the genes selected by all the three methods
is CCDC56, it has the largest Lasso and MCP estimates and is also most
significant based on SPIDR.
This gene maps to human chromosome 17q21 and  encodes the CCDC56 (coiled-coil domain containing 56) protein with 106 amino acid single-pass membranes. BRCA1 is located at 17q21-q24 and is in the neighborhood of CCDC56. Interestingly, another key tumour suppressor gene p53 also maps to chromosome 17.  


On the other hand, there are genes not selected by the Lasso or MCP
but selected by SPIDR.
An interesting one is gene UHRF1, which plays a major role in the G1/S transition and functions in the p53-dependent DNA damage checkpoint. Multiple transcript variants encoding different isoforms have been found for this gene (www.ncbi.nlm.nih.gov).
UHRF1 is a putative oncogenic factor over-expressed in several cancers, including the bladder and lung cancers. It has been reported that UHRF1 is responsible for the repression of BRCA1 gene in sporadic breast cancer through DNA methylation (Alhosin et al. (2011)). Another interesting finding based on SPIDR is a gene called SRPK1. This gene is upregulated in breast cancer and its expression level is proportional to the tumor grade.
Targeted SRPK1 treatment appears to be a promising way to enhance the effectiveness of chemotherapeutics drugs (Hayes et al. (2006, 2007)).
Other interesting findings include several genes (CDC6, CDC20, CDC25C and CDCA2) that play key roles in the regulation of cell division and interact with several proteins at multiple points in the cell cycle (www.ncbi.nlm.nih.gov).

In this example we focus on illustrating the application of SPIDR.  So we mainly highlight a few genes from the SPIDR analysis to confirm that it does reveal additional information from the data. A detailed description of the available biological functions of the selected genes is not included there, but can found from public database such as the website
of National Center for Biotechnology Information (www.ncbi.nlm.nih.gov).

\begin{figure}[H]
 \centering
 \includegraphics[width=\textwidth]{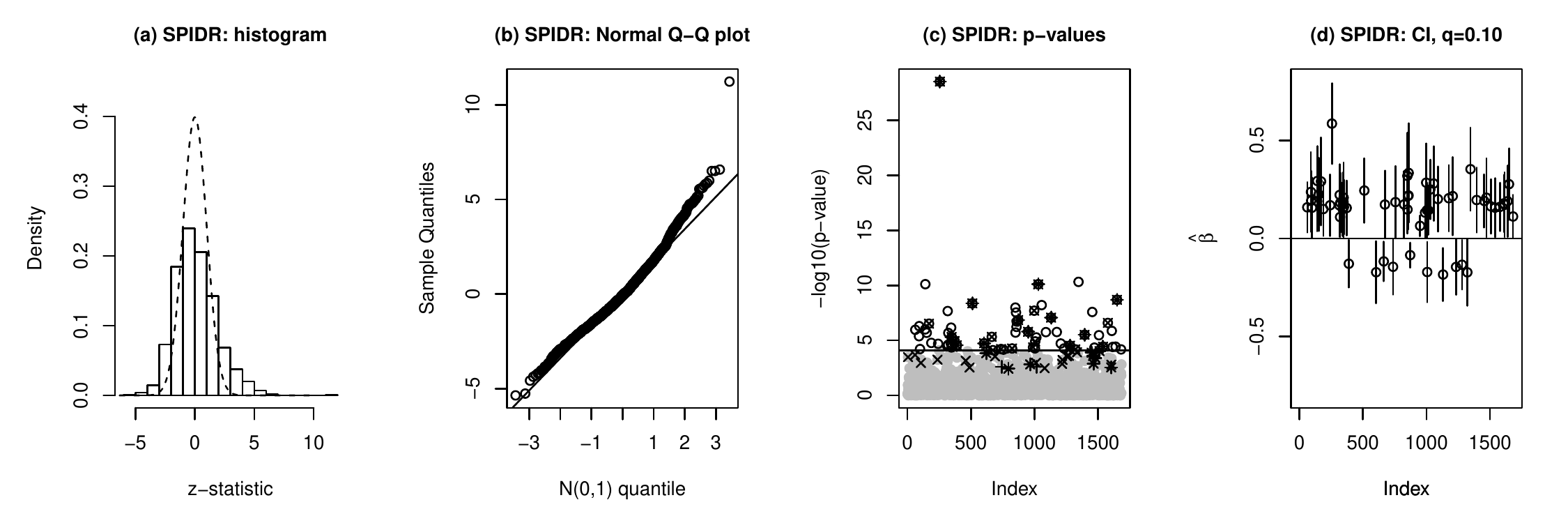}
 \caption{\label{bc-analysis2}
 Breast cancer data. (a) Histogram of SPIDR $z$-statistics, the dashed curve represents the density function of $N(0,1)$; (b) Normal Q-Q plot for the SPIDR estimates; (c) SPIDR $p$-values; (d) SPIDR confidence intervals for the selected coefficients.}
\end{figure}

In Figure \ref{bc-analysis2}, plot (a) shows the histogram of the SPIDR $z$-statistics, the dashed curve represents the standard normal density function. The distribution of the SPIDR $z$-statistics has much heavier tail than the standard normal distribution and is slightly skewed to the right. This is due to the fact that some of the $z$-statistics are not
from the null hypothesis. This can also be caused by correlation among
$z$-statistics even if their marginal distributions are $N(0, 1)$. Such
phenomenon has also been observed by Efron (2007)
in the context of detecting differentially expressed genes using microarray data. This can also be clearly seen in the normal Q-Q plot
(b). Plot (c) shows the negative $\log_{10}$ $p$-values for the SPIDR $z$-statistics. The cutoff for the negative $\log_{10}$ $p$-values for significance corresponding to FDR $q=0.1$ is 4.10, which is represented by the horizontal line in the plot. For comparison, the $p$-values of the variables selected by Lasso and MCP are also indicated in the plot
by plus ``+'' and cross ``x'' signs, respectively. Plot (d) shows the SPIDR confidence intervals for the selected coefficients.

Figure \ref{bc-analysis2} provides a panel of useful summaries of the SPIDR analysis that can be used for statistical inference, including the distribution of z-statistics, the comparison with the normal distribution
via Q-Q plot, the $p$-values and an indication of statistical significance according to a desired FDR control level, and the interval estimates of the selected effect sizes. These can be easily explained to the scientific investigators.
It is best to use Figure \ref{bc-analysis2} in combination with plots such as Figures \ref{bc-paths1} and \ref{bc-analysis1} to give a clear view of the selection results along with tuning.

\section{Discussion}

SPIDR is built on two relatively recent important developments in
high-dimensional statistics, penalized estimation and direct FDR control.
It makes the connection between these two ideas and combines them in the context of variable selection. To study the theoretical property of the proposed SPIDR estimator, we introduced the concept of an ideal estimator and provided sufficient conditions under which the SPIDR estimator is ideal with high probability.

There is a host of questions related to SPIDR that we have not been able to addressed in this paper. As we have noted based on our simulation studies, SPIDR has two interesting features that we referred to as stableness and stickiness. We considered a measure of stickiness
in Section \ref{theory}.
It would also be useful to provide a quantitative measure of stableness. Intuitively, stableness is related to the change or lack thereof in the SPIDR solution
path $\hbeta(\lam)=(\hbeta_1(\lam), \ldots, \hbeta_p(\lam))'$  with respect to $\lam$ in an appropriate interval. Therefore, in addition to the solution path itself, it would also be interesting to study the derivative of the solution path. This requires establishing the differentiability of $\hbeta(\lam)$ for $\lam$ in an interval. Whether or not this is true is not clear.
Note that $\hbeta_j(\lam)$ is perhaps not differentiable at the transition points where the MCP solution path changes direction.

The proposed method can be extended in several directions.
First, it can be applied to other regression models, including the generalized linear and Cox models. In these models, instead of using the quadratic loss in (\ref{SP1}), we can use the negative log-likelihood or partial log-likelihood as the loss functions. Of course, detailed analysis of the theoretical properties of SPIDR in these models requires further work. Second, it is possible to consider the coefficients in groups and carry out the estimation one group at a time. In particular, SPIDR can be naturally extended to group selection problems with various types of group penalties, including the group Lasso and concave group penalties. However, in group selection, the definition of FDR needs to be modified accordingly. Third, the idea of SPIDR can be applied to semiparametric and nonparametric regression models such as the partially linear and generalized additive models.

Motivated by the concept of an ideal estimator, we can also use the following two-stage approach to constructing an estimator of $\beta$. Let $\hS^*$ be the set of variables selected based on the fully penalized criterion (\ref{FP1}) with the MCP penalty. Let $\hS^*_j = \{k\in \hS^*, k\neq j\}, 1\le j\le p$. Consider the unpenalized least squares solution
\[
(\hbeta_j^*, \hbbeta_{\hS_j^*}^*) = \argmin_{\beta_j, \bbeta_{\hS_j^*}} \|\by - \bx_j \beta_j - X_{\hS_j^*}\bbeta_{\hS_j^*}\|^2, 1\le j \le p.
\]
We can use $\hbeta_j^*$ as an estimator of $\beta_j$. It can be shown that this estimator equals the ideal estimator with high-probability.
However, our simulation studies indicate that this two-stage approach
does not work as well as SPIDR,
in particular, in the presence of
strong correlation among predictors. Intuitively, this is because this two-stage method strongly depends on how well a single $\hS^*$
does as an estimator of $S$. In contrast, in SPIDR, each estimator
$\hbeta_j$ has its own estimator $\hS_j$, which tends to be better suited
for estimating $\beta_j$.
Other methods can also be considered for constructing asymptotically normal estimators in high-dimensional linear models, for example, the estimators proposed by Zhang and Zhang (2012) and Van de Geer et al. (2013) based on the efficient score approach. These estimators are computationally more demanding since they require two penalized calculations for each coefficient, one for parameter estimation and one for efficient score construction. It would be interesting to conduct
a detailed comparison of the theoretical and empirical properties of
these estimators with the proposed semi-penalized estimator, but this is beyond the scope of the present paper.

The estimation of FDR with correlated statistics is a challenging problem. In addition to the difficulty caused by correlation, false discovery proportion is inherently variable in sparse models when the number of findings is relatively small. A small change in either the number of findings or the  number of falsely selected variables can cause a big change in the proportion. We used the method of Efron (2007), which is easy to implement and computationally efficient.   Our simulation studies indicate that it can yield unbiased estimates, although the variability is relatively high. Other methods can be used in estimating the FDR in the presence of correlation, for example, the method of Fan et al. (2013). It would also be particularly interesting to develop methods tailored to
the covariance structure given in (\ref{Var1}) and (\ref{Cov1}).

In the implementation, we used the R package
\textsl{ncvreg} to compute the SPIDR solutions. It is useful to develop more efficient algorithms. Also, SPIDR appears especially suitable to be implemented in parallel, which should speed up the computation considerably.
Finally, in applications we recommend applying  SPIDR in combination with penalized selection, as illustrated in the breast cancer data example in Section \ref{NS}. In particular, it is helpful to present figures
similar to Figures \ref{bc-paths1} to \ref{bc-analysis2} to summarize
the analysis results from both penalized selection and SPIDR. Our simulation studies and data example suggest that SPIDR is a useful method for high-dimensional statistical inference in practice.

\begin{center}
{\sc Acknowledgements}
\end{center}

The research of Huang and Ma is partially supported by grants from the U.S. National Institutes of Health and National Science Foundation. The research of Zhang is partially supported by grants from the U.S. National Science Foundation and National Security Agency.
The research of Zhou is  partially
supported by the National Natural Science Funds for Distinguished Young Scholar  and Creative Research Groups of China, Shanghai University of Finance and Economics through Project 211, and Shanghai Leading Academic Discipline Project.

\section{Appendix}

\noindent
\textbf{Verification of (\ref{hbeta2}).}
The solution to (\ref{SP1}) satisfies
\begin{align*}
&X_{\hS_j}^\T (\by-\bx_j\hbeta_j-X_{\hS_j}\hbbeta_{\hS_j})
=n\drho(\hbbeta_{\hS_j};\lam),\\
&\bx_j^\T (\by-\bx_j\hbeta_j-X_{\hS_j}\hbbeta_{\hS_j})=0.
\end{align*}
The first equation gives
\(
\hbbeta_{\hS_j}=(X_{\hS_j}^\T X_{\hS_j})^{-1}X_{\hS_j}^\T (\by-\bx_j\hbeta_i)
-n(X_{\hS_j}^\T X_{\hS_j})^{-1}\drho(\hbbeta_{\hS_j};\lam).
\)
Thus
\(
X_{\hS_j}\hbbeta_{\hS_j} = P_{\hS_j}(\by-\bx_j\hbeta_j)
-X_{\hS_j}\Sigma_{\hS_j}^{-1}\drho(\hbbeta_{\hS_j};\lam).
\)
Substituting this expression into the second equation gives
\(
\bx_j^\T \{Q_{\hS_j}(\by-\bx_j\hbeta_j)+
X_{\hS_j}\Sigma_{\hS_j}^{-1}\drho(\hbbeta_{\hS_j};\lam)\}=0.
\)
It follows that
\(
\hbeta_j=(\bx_j^\T Q_{\hS_j}\bx_j)^{-1}\bx_j^\T (Q_{\hS_j}\by+
X_{\hS_j}\Sigma_{\hS_j}^{-1}\drho(\hbbeta_{\hS_j};\lam).
\)
This verifies (\ref{hbeta2}).

\medskip\noindent
\textbf{Consistency of $\hsigma^2$ in (\ref{hsigma1}).}
Let $(\by^{(1)}, X_{\hS}^{(1)})$ and $(\by^{(2)}, X_{\hS}^{(2)})$
represent the data in the partitions $D_1$ and $D_2$ with predictors in $\hS$, where $\hS$ is the
set of variables selected based on the full dataset.
For simplicity, we set $n_1=n_2=n/2$.
Then the least squares estimator based on
$(\by^{(1)}, X_{\hS}^{(1)})$ is
\[
\hbb^{(1)} = (X_{\hS}^{(1)\prime}X_{\hS}^{(1)})^{-1}X_{\hS}^{(1)\prime}\by^{(1)}.
\]
Under the conditions of Theorem \ref{Thm2}, $\hS=S$ with probability tending to 1 (Zhang (2010)).
Thus we can replace $\hS$ by $S$ in showing the
consistency here. Therefore,
since $\by^{(1)}=X_S^{(1)}\bbeta_S+\bveps^{(1)}$, we have
\begin{equation}
\label{hbb1}
\hbb^{(1)} = \bbeta_S + (X_S^{(1)\prime}X_S^{(1)})^{-1}X_S^{(1)\prime}\bveps^{(1)}.
\end{equation}
It follows that
\[
\rE\|\by^{(2)}-X_S^{(2)}\hbb^{(1)}\|^2
=\rE\|\bveps^{(2)}- X_S^{(2)}(\hbb^{(1)}-\bbeta_S)\|^2
=n_2\sigma^2 + \rE\|X_S^{(2)}(\hbb^{(1)}-\bbeta_S)\|^2.
\]
Here the cross product term vanishes because of the independence between
$\bveps^{(2)}$ and $\{X^{(2)}, \hbb^{(1)}\}$.
By (\ref{hbb1}), the independence between $X_S^{(1)}$ and $X_S^{(2)}$ and after some algebra,
\[
|S|^{-1}\rE\|X_S^{(2)}(\hbb^{(1)}-\bbeta_S)\|^2 =
\sigma^2
\trace\{(X_S^{(1)\prime}X_S^{(1)}/(n_1|S|))^{-1}
(X_S^{(2)\prime}X_S^{(2)}/(n_2|S|)\}
\to \sigma^2.
\]
Combining the above two equations we obtain
\[
(n_2+|S|)^{-1}\rE\|\by^{(2)}-X_S^{(2)}\hbb^{(1)}\|^2 \to \sigma^2.
\]
This proves the consistency of $\hsigma^2$.

We now prove Theorems \ref{Thm1} and \ref{Thm2}. The key is to show
that $\hbbeta_{-j}$ defined in (\ref{SP2a}) has the oracle properties as
the MCP solutions in Zhang (2010). Since the criteria
(\ref{SP2a}) are penalized weighted least squares,
this can be proved following the methods of Zhang (2010) with some modifications. So we only present an outline of the arguments here.

\noindent
\textbf{Proof of Theorem \ref{Thm1}.}
Let $B_j =\{\hbbeta_{-j}(\lam)\neq \tbbeta_{-j}\mbox{ or }
\sgn(\hbbeta_{-j}(\lam))\neq \sgn(\bbeta_{-j})\}$.
By the definition of $\tbbeta_{-j}$, we have
\begin{equation}
\label{OP3}
\tbbeta_{-j} = \argmin_{\bbeta_{-j}}\{ \frac{1}{2n}\|Q_j(\by-X_{-j}\bbeta_{-j}\|^2, \bbeta_{S_j^c}^o=0\}.
\end{equation}
Thus
\[
\bx_k'Q_j(\by-X_{-j}\tbbeta_{-j})=0 \text{ for } k \in S_j.
\]
Also, $\drho(\hbeta_{-j,k};\lam)=0$ if $|\hbeta_{-j,k}| \ge \gam\lam$,
where $\hbeta_{-j,k}$ is the $k$th element of $\hbbeta_j$.
Therefore, $\tbbeta_{-j}$ is a solution to (\ref{SP2}) and
$\sgn(\hbbeta_{-j})=\sgn(\bbeta_{-j})$ in the intersection of
\begin{equation}
\label{Omega1}
\Omega_{j1}(\lam) = \left\{
\max_{k \not\in S_j}|\bx_k'Q_j(\by-X_{-j}\tbbeta_{-j})|/n<\lam_1\right\}
\mbox{ and }
\Omega_{j2}(\lam) = \left\{
\min_{k \in S_j}\sgn(\beta_k)\tbeta_{-j,k} >\gam \lam\right\}.
\end{equation}
Thus
$\rP\{B_j\} \le 1-\rP\{\Omega_{j1}(\lam)\} + 1-\rP\{\Omega_{j2}(\lam)\}$.
Following the proof of Theorem 4 of Zhang (2010), we have
$\rP\{B_j\} \le 3\eps/p$.  Since $\{\hS_j\neq S_j\} \subseteq B_j$,
\[
\rP\{\cup_{j=1}^p(\hS_j\neq S_j)\} \le \sum_{j=1}^p\rP\{B_j\} \le 3 \eps.
\]
Similarly,
\[
\rP\{\cup_{j=1}^p(\hbeta_j(\lam)\neq \hbeta_j^o)\}\le \sum_{j=2}^p\rP\{B_j\} \le 3 \eps.
\]
This completes the proof.

\medskip\noindent
\textbf{Proof of Theorem \ref{Thm2}.}
For $m \ge 1$ and $B \subset \{1, \ldots, p\}\setminus\{j\}$, let
\[
\varsigma_j(\bv;m,B) = \max\left\{ \frac{\|(P_A-P_B)\bv\|}{(mn)^{1/2}}:
B\subseteq A\subseteq \{1,\ldots, p\}\setminus \{j\},
|A|=m+|B|\right\},
\]
for $\bv \in \real^n$, where $P_A$ is the orthogonal project matrix
from $\real^n$ to the linear span of $\{Q_j\bx_k: k \in A\}$.
Let $\Omega_{3j}(\lam)=\{\varsigma_j(\bveps;m^*,S_j)\le \lam\}$.
Following the proof of Theorem 5 of Zhang (2011), we have
\[
\rP\{\hbbeta_j\neq \tbbeta_j \text{ or }
\sgn(\hbbeta_j)\neq \sgn(\bbeta_j)\}
\le \sum_{k=1}^3 (1-\rP\{\Omega_{jk}(\lam)\}),
\]
where $\Omega_{jk}, k=1, 2$ are defined in (\ref{Omega1}).
This inequality and Theorem 5(ii) of Zhang (2011) imply
\(
\rP\{\hS_j\neq S_j\}\le 3\veps/p.
\)
Therefore,
\(
\rP\{\cup_{j=1}^p (\hS_j(\lam) \neq S_j)\} \le 3 \veps.
\)
Similarly, we have
\(
\rP\{\cup_{j=1}^p (\hbeta_j(\lam) \neq \tbeta_j)\} \le 3 \veps.
\)
This completes the proof.

\end{document}